\documentclass[Journal, twocolumn]{IEEEtran}

\usepackage{cite}
\usepackage{comment}
\usepackage{amsmath, amssymb, multirow, cite , epsfig, graphicx , graphics, wrapfig} 
\usepackage{color} 
\usepackage{epsfig, epstopdf}
\usepackage{graphicx , color, caption, subcaption} 

\usepackage{algorithm}
\usepackage[noend]{algpseudocode}
\makeatletter
\def\BState{\State\hskip-\ALG@thistlm}
\makeatother

\begin{document} 
\title {Optimal Measurement Policy for Predicting UAV Network Topology}

\author{
\IEEEauthorblockN{Abolfazl Razi \IEEEauthorrefmark{1}, Fatemeh Afghah \IEEEauthorrefmark{1}, Jacob Chakareski \IEEEauthorrefmark{2}\\
\IEEEauthorblockA{\IEEEauthorrefmark{1}School of Informatics, Computing and Cyber Systems, Northern Arizona University, Flagstaff, AZ}\\
\IEEEauthorblockA{\IEEEauthorrefmark{2}Departmnt of Electrical and Computer Engineering, University of Alabama, Tuscaloosa, AL}\\
}
}

\maketitle

\begin{abstract}

In recent years, there has been a growing interest in using networks of Unmanned Aerial Vehicles (UAV) that collectively perform complex tasks for diverse applications.  
An important challenge in realizing UAV networks is the need for a communication platform that accommodates rapid network topology changes. For instance, a timely prediction of network topology changes can reduce communication link loss rate by setting up links with prolonged connectivity.

In this work, we develop an optimal tracking policy for each UAV to perceive its surrounding network configuration in order to facilitate more efficient communication protocols. More specifically, we develop an algorithm based on particle swarm optimization and Kalman filtering with intermittent observations to find a set of optimal tracking policies for each UAV under time-varying channel qualities and constrained tracking resources such that the overall network estimation error is minimized. \end{abstract}


\section{Introduction:} 
Utilizing a swarm of autonomous UAVs to perform complicated tasks in military and commercial applications has recently gained a lot of momentum and is expected to continue growing in the coming years \cite{UAV_Roadmap1}. 
For instance, the U.S. Navy has recently launched a prototype for the LOw-Cost Unmanned aerial vehicle Swarming Technology (LOCUST) project that implements the required technology for UAV swarm attacks~\cite{LOCUST1}.
 The UAV market value was estimated to be USD 13.22 Billion in 2016 and is projected to reach USD 28.27 Billion by 2022 \cite{UAV_Market1}.

Swarms of fast-flying UAVs form dynamic networks with rapidly changing topologies, where conventional communication protocols that make decisions at different layers solely based on current network configuration and ignore topology evolution often perform poorly \cite{Razi_TCCN,2012_RaziCISS}. For instance, a conventional relay selection algorithm, which selects intermediate relay nodes merely based on the current system configuration falls behind the abrupt topology changes and hence fails in providing prolonged network connectivity for member UAVs \cite{FANETSurvey2013}. 
As such, developing algorithms that enable the UAVs to effectively predict their surrounding nodes' motion trajectories can significantly improve the performance of topology-aware communication and control algorithms.

\begin{figure}[t]
\centering
\includegraphics[width=.9\linewidth]{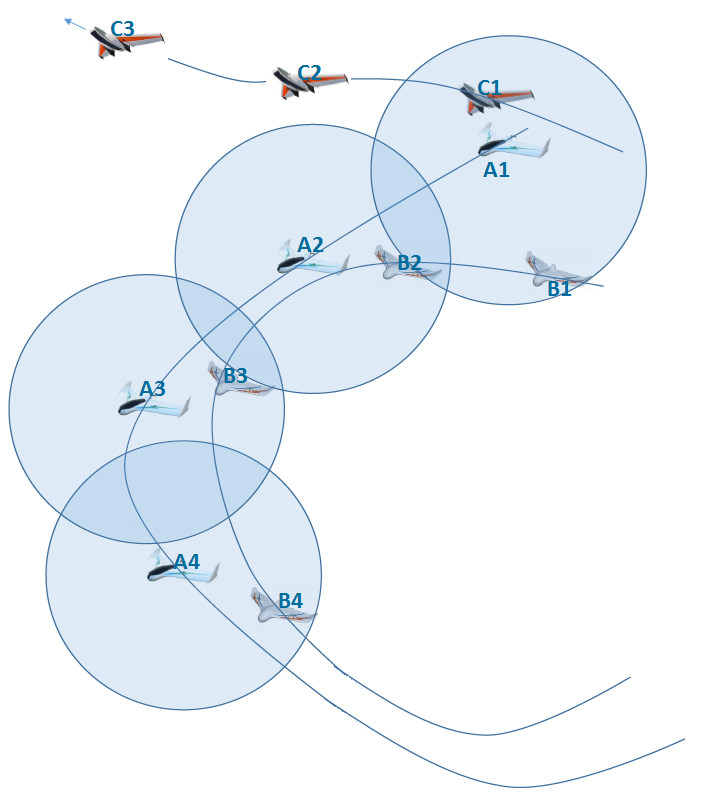}
\caption{\small Motion trajectory of 3 UAVs (A,B and C).} 
\label{fig:connectivity}
\vspace{-15pt}
\end{figure}

An example for the utility of predicting network topology in optimizing communication protocols is illustrated in Fig. \ref{fig:connectivity}.

The motion trajectory of 3 UAVs ($A, B, C$) are depicted in figure 2. UAV1 has limited communications range that is shown by three circles centered at locations ($A1, A2$ and $A3$) for three time points $t_1,t_2,t_3$. In order to retain connectivity with at least one neighboring UAV, $A$ intends to choose the best relay node. At time $t_1$, $B$ and $C$ are both within the $A$'s accessible range. Under a conventional relay selection method, $A$ chooses $B$ since it is closer in distance and requires less transmission power (or yields lower delay). At time $t_2$, $B$ moves out of $A$'s accessible range and hence $A$ needs to hand over to another relay node, which requires additional signaling. Also, if node $C$ has already been assigned with a relaying task for another UAV, then $A$ falls out of the network and looses its connection to the base station. However, if $A$ was capable of predicting motion trajectory of neighbor UAVs and recognizing a better alignment between $C$'s and its own trajectories, it would have chosen $C$ at time $t_1$, which never goes out of the accessible range. Consequently, $B$ would have retained the connectivity until the end of its mission \cite{2017_UAV_WiSEE}.

In this paper, we use the popular method of Dubin's curves to model the motion of UAVs and use the time and measurement update equations developed for Kalman filtering with intermittent observations to predict the locations of UAVs, when the tracking measurements are sporadically available. In order to obtain the next network prediction results and retain maximal connectivity \cite{ZhangH05_K_connectivity, WSN2008connectivity}, we propose an optimal measurement policy for UAVs such that the prediction error of the surrounding UAV locations are minimized under a constrained measurement resource and an interrupted observation model. The idea is to use particle swarm optimization to develop a resource management policy for an individual UAV to allocate its measurement resources optimally among a set of $k$ neighbor UAVs such that the expected error covariance is minimized. The proposed method provides a practical solution for a theoretically intractable problem and has the advantage of adapting to time-varying measurement conditions.

\section{Topology Prediction}
Recently, several methods including data-driven methods \cite{Lee2015trajectory1, Ossama201145}, piecewise segment methods \cite{Choi2006trajectory}, hidden Markov models \cite{Bennewitz05learningmotion}, Levy and Levy flight process \cite{gonzalez08_peoplePattern}, manifold learning \cite{Lee2012Manifold}, and Gaussian mixture models \cite{Aoude2011BayesianTrajectories} are proposed to model mobile user motions (e.g. vehicles and pedestrians) in wireless networks.
However, these models are not well suited for freely flying UAVs that do not follow man-made or natural path profiles (e.g. roads and streets). Here, we use the Dubin's curve method, which is general and specifies the motion trajectory of a moving object based on the exerted forces in 3 dimensions as follows: \cite{Lin2014DubinCurve_UAV}.

\begin{align} \label{eq:model}
\begin{cases}
\mathbf{x}_i[k+1]&= A \mathbf{x}_i[k] + B \mathbf{u}_i[k] + \mathbf{w}_i[k],\\
\mathbf{y}_i[k]&=\gamma_i[k] C\mathbf{x}_i[k]+ \mathbf{v}_i[k], 
\end{cases}
\vspace{-5pt}
\end{align}
where we have:
\vspace{-5pt}
\begin{align}
\nonumber
&\mathbf{x}_i[k] = \begin{bmatrix}x_i[k] ~ y_i[k] ~ z_i[k] ~ v_{xi}[k] ~  v_{yi}[k]~ v_{zi}[k] \end{bmatrix}^T, ~\text{(state vector)} \\
\nonumber
&\mathbf{u}_i[k] = \begin{bmatrix}a_{xi}[k] ~ a_{yi}[k] ~ a_{zi}[k] \end{bmatrix}^T,~~~\text{(input vector)}\\
\nonumber
&\mathbf{y}_i[k]=\begin{bmatrix} y_{xi}[k]~ y_{yi}[k] ~ y_{zi}[k] \end{bmatrix}^T, ~~~\text{(observation vector)}\\
\nonumber
&A = \begin{bmatrix}  
\mathbf{I}_{3 \times 3}  & dt\mathbf{I}_{3 \times 3}\\
\mathbf{0}_{3 \times 3} & \mathbf{I}_{3 \times 3}
\end{bmatrix},
B= \begin{bmatrix}  
\mathbf{0}_{3 \times 3} & \mathbf{I}_{3 \times 3}\end{bmatrix},
C=\begin{bmatrix} \mathbf{I}_{3 \times 3} & \mathbf{0}_{3 \times 3} \end{bmatrix}^T,\\
&\mathbf{v}_i[k]\sim \mathcal{N}(\mathbf{0}, \mathbf{R}_i),~~  \mathbf{w}_i[k]\sim \mathcal{N}(\mathbf{0}, \mathbf{Q}_i).
\end{align}

Here, dt is the time step of the discretized system, $\mathbf{x}_i[k]$ is the state vector of UAV $i$ at time $k$, representing the location ($x,y,z$) and velocities ($v_x,v_y,v_z$), $\mathbf{u}[k]$ captures the impact of applied forces on the accelerations, $\mathbf{y}[k]$ is the measurements provided by the tracking system. We assume that $\mathbf{w}$ and $\mathbf{v}$ are zero mean Gaussian random vectors with covariances $\mathbf{Q}$ and $\mathbf{R}$ which respectively represent the model turbulence and observations noise. 
The measurements success is modeled as a sequence of Bernoulli distributed random variables ($\gamma[k] \in \{0,1\}, Pr(\gamma[k]=1)=\lambda$)~\cite{Sinopoli2003IntermittentKalman}. 
It is known that an optimal estimation of the state vectors (in MMSE sense) is obtained using Kalman filtering with the following steps:
\begin{align}
\label{eq:kalman}
\nonumber
&\begin{cases}
\tilde{\mathbf{x}}[k]=A \hat{\mathbf{x}}[k-1] + B\mathbf{u}[k-1] \\
\tilde{P}[k]=A \hat{P}[k-1]A^T + Q 
\end{cases} (\text{time update eqs}),~~~~\\
&\begin{cases}
K[k]= \tilde{P}[k]C^T(C\tilde{P}[k]C^T+R)^{-1} \\
\hat{x}[k]=\tilde{x}[k]+K[k] (y[k]-C\tilde{x}[k]), \\
\hat{P}[k]= (I-K[k]C)\tilde{P}[k]
\end{cases} (\text{measurement eqs}),
\end{align} 
where the measurement update equations are performed only for $\gamma[k]=1$ and we set $\hat{x}[k]=\tilde{x}[k], \hat{P}[k]=\tilde{P}[k]$ for $\gamma[k]=0$ \cite{Sinopoli2003IntermittentKalman}. 
Under some mild convergence conditions on the A,~B,~C,~Q matrices, (i.e. observability of ($A,C$), controllability of ($A,B$), stabilizability of ($A,Q^{1/2}$), and bounded system and measurement noise covariances $tr(Q)<\infty,tr(R)<\infty$), the sequential error covariance matrices $\hat{P}[k]$ starting from any initial value $\hat{P}[0]$ converge to a unique limit, which is the solution of the following Modified Discrete-time Algebraic \textit{Ricatti} equations (MARE)~\cite{Sinopoli2003IntermittentKalman}: 
\begin{align}
\label{eq:MARE}
\nonumber
&g(P; \lambda,R,Q)=APA^T+Q- \lambda APC^T(CPC^T+R)^{-1}CPA^T\\
&P=g_{\lambda,R,Q}(P)
\end{align}

The solution of (\ref{eq:MARE}) for intermittent observation is not known in general, but its statistical properties are studied under different assumptions \cite{KalmanIntermittentgame_2013, KalmanIntermittent_2010 ,KalmanIntermittent_2011, KalmanIntermittent_2012, KalmanIntermittent_2014, KalmanIntermittent_2015, KalmanIntermittent_2016}.
In particular, it is known that the convergence of $\hat{P}[k]$ is ensured if the observation availability occurs with probability $\lambda \geq \lambda_c$, where $\lambda_c$ is a critical value with known upper and lower bounds \cite{KalmanIntermittent_2015, Sinopoli2004IntermittentKalman}.

\begin{figure}[t] 
    \centering
    \begin{subfigure}[h]{0.9\linewidth}
	\label{fig:2-1}
        \centering
        \includegraphics[width=.9\textwidth]{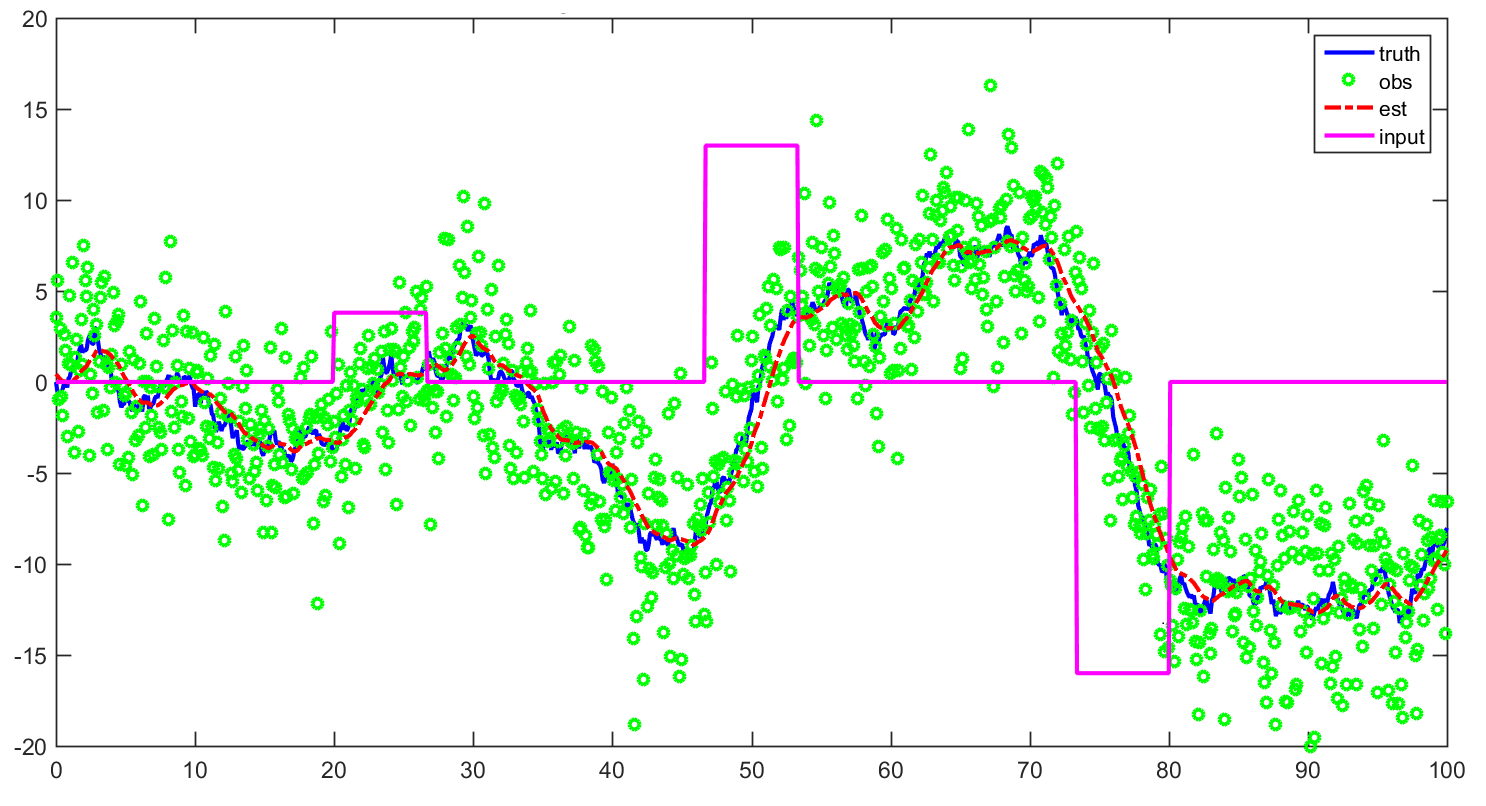} 
        \caption{\small observation rate: $\lambda =1$}
    \end{subfigure}%
\\
    \begin{subfigure}[h]{0.9\linewidth}
	\label{fig:2-2}
        \centering
        \includegraphics[width=.9\textwidth]{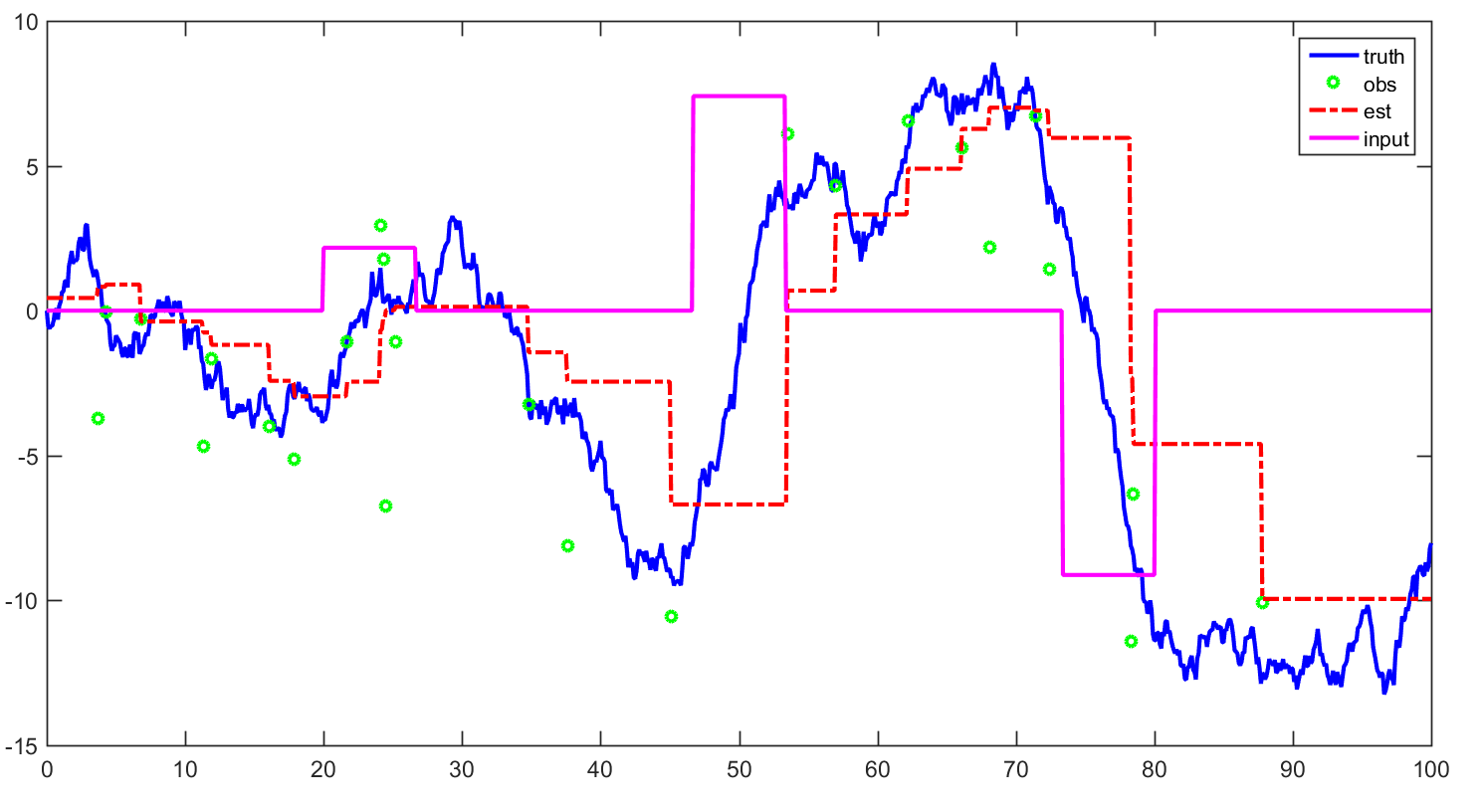}
        \caption{\small observation rate: $\lambda =0.02$}
        \end{subfigure}
\caption{\small prediction result of a 1-D motion using Kalman filtering with intermittent observation. The parameters are $dt=0.1, Q=0.1I_n, R=10I_m, a[k]\sim \mathcal{N}(\mathbf{0}, 100\mathbf{I})$. The resulting estimation error is MSE: $0.048$ for $\lambda=1$ and MSE: $0.366$ for $\lambda=0.2$.}
\label{fig:intermittent}
\vspace{-15pt}
\end{figure}

Here, we assume that the input vector, $\mathbf{u}$ is known. Note that the role of unknown input and the time update equations in general is more important for intermittent observations, since we rely more on the time-update equations. Intuitively, in a fully observable system, the measurement update equations enhance our estimates and partially compensate the lack of information about the unknown inputs, as expected. Fig.~\ref{fig:intermittent} demonstrates the impact of unknown input for a 1D motion. 
However, we can use recently proposed techniques to tackle the case of unknown input vectors. For instance, with a simple conversion technique, the unknown input can be absorbed into the state vector to transform the system to an equivalent system with known inputs \cite{Mohamed2007UnknownKalmanInput}. If the statistics of the unknown input is fully known, then the optimal state estimator can be represented by the two-stage Kalman filter \cite{Hsieh:2011:Kalman2UnInput}.
The simulation results for the case of unknown input and $\lambda=1$ and $\lambda=0.2$ is shown in Fig.\ref{fig:intermittent}.

\section{Optimal Tracking Policy}  \label{sec:scenario}
In this section, we assume a scenario, where a UAV intends to track $N$ surrounding UAVs. The parameters for target UAV $i$ includes the motion control noise covariance $R_i$, navigation noise covariance $Q_i$, and the packet transmission success rate $p(\gamma_i=1)=\lambda_i$. The observer UAV is assumed to be equipped with $M$ tracking instruments, where $\rho=M/N < 1$. The objective is to design an algorithm to optimally assigns the tracking resources among the $N$ surrounding UAVs, during a measurement cycle composed of $T$ consecutive time slots, such that a desired evaluation metric is minimized. We develop the algorithm in two sequential steps for a measurement cycle. 
We first consider a probabilistic model, where at each time slot, UAV $i$ is tracked with probability $\alpha_i$, where $0\leq \alpha_i \leq 1$. The objective of this step is to find the \textit{optimal measurement probability vector} $\mathbf{\alpha}=[\alpha_1,\alpha_2,\dots,\alpha_N]^T$, such that a desired performance criterion is met. 
Apparently, due to resource constraint, we have $|\mathbf{\alpha}|_1= \alpha_1+\alpha_2+\dots \alpha_N \leq M=N\rho$. We consider the following two objective functions.\\
One may choose to minimize the worst-case squared error by solving the following optimization problem: 
\begin{align}
\nonumber
(\alpha_1^*, \alpha_2^*, \dots , \alpha_N^*)&=\underset{(\alpha_1, \alpha_2, \dots , \alpha_N)}{\text{minimize}}{ \max_{i\in\{1,2,\dots,N\}}E[tr(\hat{P}_i[k])] }\\
\nonumber&\text{subject to:  } \alpha_i \geq 0,~~~~~~ \text{for } i=1,2,\dots,N\\
\nonumber&~~~~~~~~~~~~~     \alpha_i \lambda_i \geq \lambda_i^{(c)}\\
&~~~~~~~~~~~~~     \alpha_1+\alpha_2+\dots \alpha_N \leq \rho N,
\end{align}
where $\lambda_i^{(c)}$ is the critical value for the channel success probability to have bounded expected error covariance, characterized in \cite{Sinopoli2003IntermittentKalman}. The feasible set of this problem is defined by its constraint functions and the solution for this problem exists only if we have sufficient tracking resources (i.e. $M \geq \sum_{i=1}^N \alpha_i \geq \sum_{i=1}^N  \frac{\lambda_i^{(c)}}{ \lambda_i }$). 
In particular, we use a water filling approach to assign sufficient tracking attempt probability $\alpha_i$ for each UAV such that its effective measurement rate $\alpha_i \lambda_i$ exceeds the corresponding critical value until all constraints are satisfied. However, finding the optimal way to associate the remaining tracking resources to optimize the objective function is not tractable.  
An alternative objective function is to minimize the overall aggregated estimation error. More formally, one may intend to solve the following optimization problem:
\begin{align}
\nonumber
(\alpha_1^*, \alpha_2^*, \dots , \alpha_N^*)&=\text{argmin}\sum_{i=1}^{N}\sum_{k=1}^{T}tr(\hat{P}_i[k])\\
\nonumber
&\text{subject to:  } \alpha_i \geq 0,~~~~~~ \text{for } i=1,2,\dots,N\\
\nonumber
&~~~~~~~~~~~~~     \alpha_i \lambda_i \geq \lambda_i^{(c)},\\
&~~~~~~~~~~~~~     \alpha_1+\alpha_2+\dots \alpha_N \leq \rho N.
\end{align}

This is a stochastic optimization problem. It is practically hard to solve due to the randomness of $\hat{P}_i[k]$, therefore we solve the following deterministic by almost equivalent problem:
\begin{align}
\nonumber
(\alpha_1^*, \alpha_2^*, \dots , \alpha_N^*)&=\text{argmin}\sum_{i=1}^{N} \lim_{k \to \infty} E[tr(\hat{P}_i[k])]\\
\nonumber
&\text{subject to:  } 0 \leq \alpha_i \leq 1,\\
\nonumber
&~~~~~~~~~~~~~     \alpha_i \lambda_i \geq \lambda_i^{(c)}\\
&~~~~~~~~~~~~~     \alpha_1+\alpha_2+\dots \alpha_N \leq \rho N
\end{align}
Note that this problem is deterministic due to the expectation operator. However, $\lambda_c$ and $\hat{P}[k]$ do not admit closed-from equations. Furthermore, estimating $\mathbf{R}[k]$ and $\mathbf{Q}[k]$ in practice requires a large number of consecutive measurement samples, which is not feasible in time-varying conditions. Therefore, finding an analytical solution for this optimization problem is not practically feasible. Therefore, we propose to use the following approximate optimization algorithm based on Particle Swarm Optimization (PSO) to solve this problem.

The main idea here is that we generate a set of $P$ particles, each representing the system with a set of hypothetical parameters $\lambda_i^{(p)}[0],\mathbf{R}^{(p)}_i[0], \mathbf{Q}^{(p)}_i[0], \hat{P}^{(p)}_i[0]$, which are randomly generated. Also, we assume a random measurement policy vector $\mathbf{\alpha}^{(p)}=[\alpha_1^{(p)},\dots,\alpha_N^{(p)}]$ to each particle $p$. Then, for each UAV $i$, we update the Kalman filtering equations and estimate the state vector $\hat{x}_i[k]$ and the estimation covariance matrix $\hat{P}_i[k]$. Then, for each particle, we use MARE to generate the current hypothetical error covariance matrix $\hat{P}^{(p)}_i[k],1=1,2,\dots,N$ based on the previous error covariance matrix $\hat{P}^{(p)}_i[k-1]$. Over time, the hypothetical error covariance matrix of a particle which has more accurate presumed parameters is expected to be closer to the estimate error covariance matrix $\hat{P}_i[k]$. As such, we move each particle in a direction which is the linear combination of three directions including: i) their motion direction in the previous iteration, ii) their motion towards a point in their local history that yields the best match (i.e. local best), and iii) towards the particle with the best match (global best). At each iteration, the parameters of each particle are updated based on the obtained particle velocity. After a number of iterations, the particles are expected to converge to a particle with the best value for observation probability vector $\mathbf{\alpha}^{(p)}$, which in turn yields the best prediction results. A formal algorithmic description of the proposed algorithm is included in Algorithm 1.

\begin{algorithm}
\caption{Optimized measurement policy design for UAV swarms using PSO}\label{algo}
\begin{algorithmic}[1]
\BState \textbf{Initialization:}\For {swarm particles p=1 to P}
\State ~Initialize measurement parameters for all UAVs ($\lambda_i^{(p)}[0],\mathbf{R}^{(p)}_i[0], \mathbf{Q}^{(p)}_i[0], \hat{P}^{(p)}_i[0], \alpha_i[k]$) with random values.  
\State set $k \gets 1$
\EndFor

\BState \textbf{top}:
\For {UAV i=1 to N}
\State Estimate $\hat{x}_i[k]$ and $\hat{P}_i[k]$ using (\ref{eq:kalman})
\EndFor
\For {swarm particles p=1 to P}
\For {UAV i=1 to N}
\State $\hat{P}^{(p)}_i[k] \gets g( \hat{P}^{(p)}_i[k-1];\lambda_i^{(p)}[k],R_i^{(p)}[k],Q_i^{(p)}[k]) $
\EndFor
\State $f(p)=\sqrt{\sum_{i=1}^N \big(\text{trace}(\hat{P}^{(p)}_i[k])-\text{trace}(\hat{P}_i[k])\big)^2}$
\If {$f(p)<f_\text{min}(p)$}                {$f_\text{min}(p) \gets f(p), bt(p) \gets k$} \EndIf
\If {$f_\text{min}(p)<f_\text{min}$}                 {$f_\text{min} \gets f_\text{min}(p), bp \gets p $} \EndIf
\If {$f_\text{min}(p)<\text{Threshold}$}                \textbf{Exit} \EndIf
\State \textbf{update swarm measurement probabilities}:
\State $\alpha_i^{(p)}[k] \gets \alpha_i^{(p)}[k-1]+\beta_L (\alpha_i^{(p)}[k-1]-\alpha_i^{(p)}[bt(k)]) + \beta_G(\alpha_i^{(p)}[k-1]-\alpha_i^{(bp)}[bt(k)])$
\State Apply the same updates for $\lambda^{(p)}_i, \mathbf{R}^{(p)}_i[k]$ and $\mathbf{Q}^{(p)}_i[k]$
\EndFor
\State $k+1 \gets k$
\State \textbf{goto} \emph{top}.
\end{algorithmic}
\end{algorithm}

Here, $\beta_L$ and $\beta_G$ are tuning parameters to balance between the motion velocities of each particle towards its local minimum and the global minimum. Also, $bp$ and $bt(p)$, respectively, store the best particle ID (global optimum) and the time index of the local optimum for particle $p$. 

Finally, we note that finding the optimal observation probability vector $\mathbf{\alpha}$ does not complete the problem. It only determines the rate under which each UAV should be subject to tracking. However, actual observation policy comprises fully determining a binary matrix $\mathbf{B}=[\beta_{it}]_{N \times T}$, where $\beta_{it}=1$ represents a measurement attempt to track UAV $i$ at time $t$. In order to determine the actual measurement attempt matrix $\mathbf{B}$, in this case, we recall the following constraints:
\begin{align}
\begin{cases}
\sum_{i=1}^N \beta_{it} < \rho N& \text{for all time slots } t=1,2,\dots,T\\ 
\frac{1}{T}\sum_{t=1}^T \beta_{it} = \alpha_i^*  &\text{for all UAVs } i=1,2,\dots,N. 
\end{cases}
\end{align}

We conjecture that one guideline to minimize the accumulated estimation error ($\sum_{t=1}^T P_i[t]$) for each UAV, is to use the most alternating measurement pattern such that measurement resources are evenly distributed over time. For instance, if $\alpha_i^*=1/3$, the optimal measurement pattern is $[\beta_{i1},\beta_{i2},\dots,\beta_{iT}]=[1001001\dots100]$. This fact is confirmed by simulation results and we are working to provide an analytical proof.

\section{Simulation Results}

\begin{figure}[h]
\includegraphics[width=0.95\linewidth]{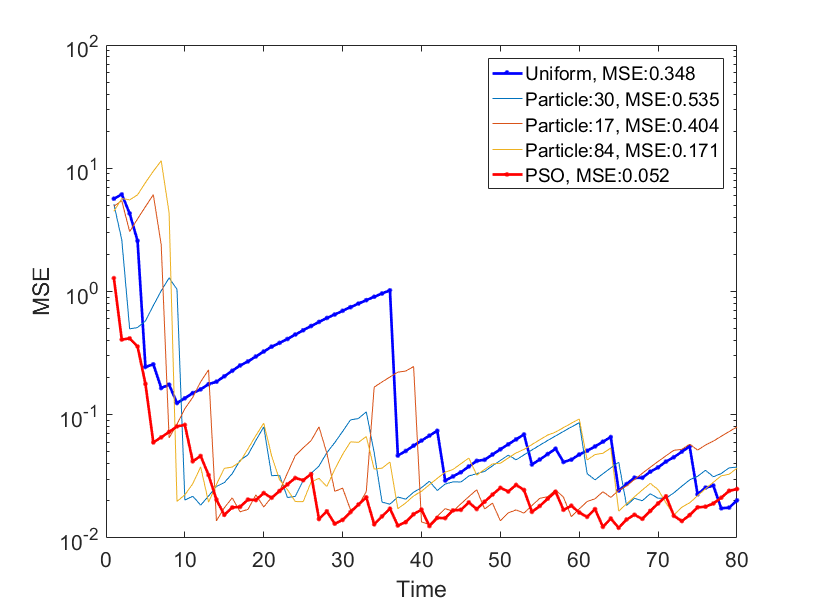} 
\caption{\small Average tracking estimation error using \textit{uniform} distribution, \textit{random} probabilities for randomly selected particles, and optimized probabilities using PSO.}
\label{fig:MSE}
\vspace{-15pt}
\end{figure}

\begin{figure}[t]
\includegraphics[width=0.95\linewidth]{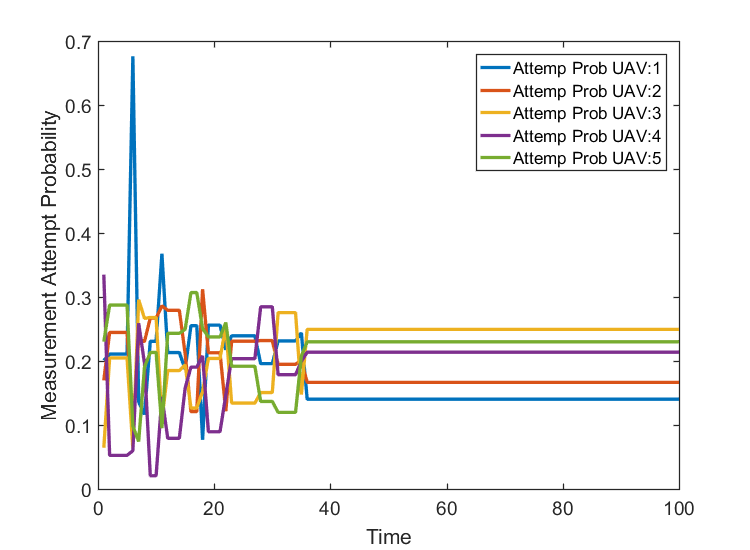}  
\caption{\small Tracking Policies ($\alpha_i^{(p)}$) corresponding to the best particle obtained using the proposed algorithm.}
\label{fig:Prob}
\vspace{-15pt}
\end{figure}

In this section, simulation results are provided to verify the performance of the proposed PSO-based algorithm. 
Fig. \ref{fig:MSE} presents the average estimation error in terms of Mean Squared Error $(MSE(k)=\frac{\sum_{i=1}^N|\mathbf{x}_i[k]-\hat{\mathbf{x}}_i[k]|_2^2}{N})$ for $N=5, M=1$ using three methods. The \textit{uniform} distribution is corresponding to the case of evenly assigning $M$ tracking resources among $N$ UAVs at each time slot (i.e. with Prob. $\alpha_i=M/N$ for $i=1,2,\dots,N$). PSO denotes the average MSE obtained by probabilities determined by the proposed PSO-based algorithm (see Algorithm 1). The MSE for $3$ random particles are also shown for the sake of comparison completeness. It can be seen that the proposed PSO algorithm outperforms the uniform and randomly selected measurement policies. The fluctuations of MSE are corresponding to alternating measurement success and fail events ($\alpha_i[k]\gamma_i[k]$). Fig. \ref{fig:Prob} also represents the measurement probability evolution by the PSO algorithm confirming that after about 40 iterations the probabilities converge to their optimal values.

\begin{figure}[t]
\includegraphics[width=0.95\linewidth]{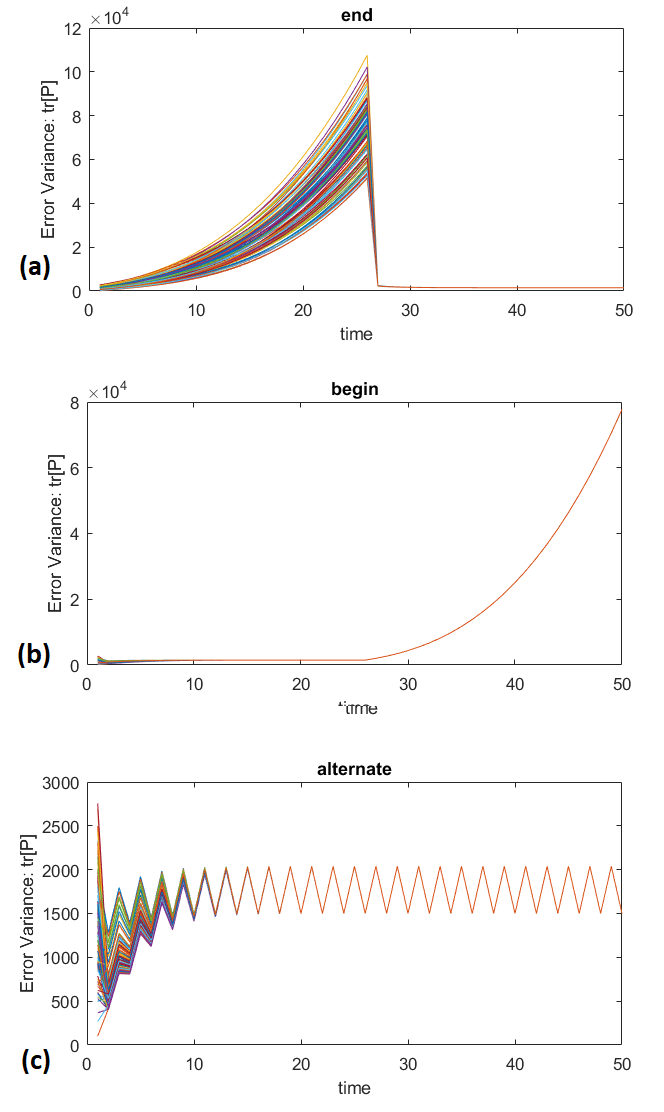}  
\caption{\small Comparison of measurement patterns for $\alpha=1/2$ for 100 scenarios with different initial error covariance matrix $P_i[0]$. (a) All $\alpha T=T/2$ measurement resources are assigned in the beginning time slots. (b) All measurement resources are assigned in the last $T/2$ time slots.(a) Measurement resources are assigned evenly over time.}
\label{fig:pattern}
\vspace{-1pt}
\end{figure}

Finally, different measurement patterns are compared in Fig. \ref{fig:pattern}, which verifies our conjecture about the optimality of the most evenly distributed patterns. In other words, it is desired that the ``1''s of each row of the actual measurement attempt matrix $\mathbf{B}=[\beta_{it}]$ are evenly distributed. The intuitive justification behind this fact is that consecutive time slots without measurements cause dramatic uncontrolled rises in estimation error, whereas as a series of consecutive measurements does not provide the same amount of improvement. As such, the most alternating pattern provides the lowest accumulated estimation error on average.

\section{CONCLUSIONS} \label{sec:conclusion}
The proposed methodology provides a numerical solution for an otherwise intractable optimization problem of assigning limited measurement resources among a large number of targets. Our approach suggests a low-complexity algorithm to optimally assign tracking resources by a UAV to discover and predict its surrounding UAVs. The proposed method is general and applicable to similar problems that use Kalman filtering as their optimization methods. 

\bibliographystyle{IEEEtran}   
\bibliography{ref}

\end{document}